\documentclass[floatfix,twocolumn,showpacs,preprintnumbers,amsmath,amssymb,prc,superscriptaddress]{revtex4} 
\usepackage{graphicx}
\usepackage{dcolumn}
\usepackage{bm}
\newcommand{\bef}{\begin{figure}[!htb]}
\newcommand{\eef}{\end{figure}}

\def \lt {\mbox{$\ <\ $}}

\usepackage{color}

\begin{document}
\title{Directed and elliptic flow of charged particles in Cu+Cu collisions at $\sqrt{\bm {s_{NN}}} =$ 22.4 GeV}

\affiliation{Argonne National Laboratory, Argonne, Illinois 60439, USA}
\affiliation{Brookhaven National Laboratory, Upton, New York 11973, USA}
\affiliation{University of California, Berkeley, California 94720, USA}
\affiliation{University of California, Davis, California 95616, USA}
\affiliation{University of California, Los Angeles, California 90095, USA}
\affiliation{Universidade Estadual de Campinas, Sao Paulo, Brazil}
\affiliation{University of Illinois at Chicago, Chicago, Illinois 60607, USA}
\affiliation{Creighton University, Omaha, Nebraska 68178, USA}
\affiliation{Czech Technical University in Prague, FNSPE, Prague, 115 19, Czech Republic}
\affiliation{Nuclear Physics Institute AS CR, 250 68 \v{R}e\v{z}/Prague, Czech Republic}
\affiliation{University of Frankfurt, Frankfurt, Germany}
\affiliation{Institute of Physics, Bhubaneswar 751005, India}
\affiliation{Indian Institute of Technology, Mumbai, India}
\affiliation{Indiana University, Bloomington, Indiana 47408, USA}
\affiliation{Alikhanov Institute for Theoretical and Experimental Physics, Moscow, Russia}
\affiliation{University of Jammu, Jammu 180001, India}
\affiliation{Joint Institute for Nuclear Research, Dubna, 141 980, Russia}
\affiliation{Kent State University, Kent, Ohio 44242, USA}
\affiliation{University of Kentucky, Lexington, Kentucky, 40506-0055, USA}
\affiliation{Institute of Modern Physics, Lanzhou, China}
\affiliation{Lawrence Berkeley National Laboratory, Berkeley, California 94720, USA}
\affiliation{Massachusetts Institute of Technology, Cambridge, MA 02139-4307, USA}
\affiliation{Max-Planck-Institut f\"ur Physik, Munich, Germany}
\affiliation{Michigan State University, East Lansing, Michigan 48824, USA}
\affiliation{Moscow Engineering Physics Institute, Moscow Russia}
\affiliation{NIKHEF and Utrecht University, Amsterdam, The Netherlands}
\affiliation{Ohio State University, Columbus, Ohio 43210, USA}
\affiliation{Old Dominion University, Norfolk, VA, 23529, USA}
\affiliation{Panjab University, Chandigarh 160014, India}
\affiliation{Pennsylvania State University, University Park, Pennsylvania 16802, USA}
\affiliation{Institute of High Energy Physics, Protvino, Russia}
\affiliation{Purdue University, West Lafayette, Indiana 47907, USA}
\affiliation{Pusan National University, Pusan, Republic of Korea}
\affiliation{University of Rajasthan, Jaipur 302004, India}
\affiliation{Rice University, Houston, Texas 77251, USA}
\affiliation{Universidade de Sao Paulo, Sao Paulo, Brazil}
\affiliation{University of Science \& Technology of China, Hefei 230026, China}
\affiliation{Shandong University, Jinan, Shandong 250100, China}
\affiliation{Shanghai Institute of Applied Physics, Shanghai 201800, China}
\affiliation{SUBATECH, Nantes, France}
\affiliation{Texas A\&M University, College Station, Texas 77843, USA}
\affiliation{University of Texas, Austin, Texas 78712, USA}
\affiliation{University of Houston, Houston, TX, 77204, USA}
\affiliation{Tsinghua University, Beijing 100084, China}
\affiliation{United States Naval Academy, Annapolis, MD 21402, USA}
\affiliation{Valparaiso University, Valparaiso, Indiana 46383, USA}
\affiliation{Variable Energy Cyclotron Centre, Kolkata 700064, India}
\affiliation{Warsaw University of Technology, Warsaw, Poland}
\affiliation{University of Washington, Seattle, Washington 98195, USA}
\affiliation{Wayne State University, Detroit, Michigan 48201, USA}
\affiliation{Institute of Particle Physics, CCNU (HZNU), Wuhan 430079, China}
\affiliation{Yale University, New Haven, Connecticut 06520, USA}
\affiliation{University of Zagreb, Zagreb, HR-10002, Croatia}

\author{G.~Agakishiev}\affiliation{Joint Institute for Nuclear Research, Dubna, 141 980, Russia}
\author{M.~M.~Aggarwal}\affiliation{Panjab University, Chandigarh 160014, India}
\author{Z.~Ahammed}\affiliation{Variable Energy Cyclotron Centre, Kolkata 700064, India}
\author{A.~V.~Alakhverdyants}\affiliation{Joint Institute for Nuclear Research, Dubna, 141 980, Russia}
\author{I.~Alekseev~~}\affiliation{Alikhanov Institute for Theoretical and Experimental Physics, Moscow, Russia}
\author{J.~Alford}\affiliation{Kent State University, Kent, Ohio 44242, USA}
\author{B.~D.~Anderson}\affiliation{Kent State University, Kent, Ohio 44242, USA}
\author{C.~D.~Anson}\affiliation{Ohio State University, Columbus, Ohio 43210, USA}
\author{D.~Arkhipkin}\affiliation{Brookhaven National Laboratory, Upton, New York 11973, USA}
\author{G.~S.~Averichev}\affiliation{Joint Institute for Nuclear Research, Dubna, 141 980, Russia}
% ...
\collaboration{STAR Collaboration}\noaffiliation
\begin{abstract}
 This paper reports results for directed flow $v_{1}$ and elliptic flow $v_{2}$ of charged particles in Cu+Cu collisions at $\sqrt{s_{NN}}= 22.4$ GeV at the Relativistic Heavy Ion Collider. The measurements are for the $0-60\%$ most central collisions, using charged particles observed in the STAR detector. Our measurements extend to 22.4 GeV Cu+Cu collisions the prior observation that $v_1$ is independent of the system size at 62.4 and 200 GeV, and also extend the scaling of $v_1$ with $\eta/y_{\rm beam}$ to this system.  The measured $v_2(p_T)$ in Cu+Cu collisions is similar for $\sqrt{s_{NN}}$ throughout the range 22.4 to 200 GeV.  We also report a comparison with results from transport model (UrQMD and AMPT) calculations. The model results do not agree quantitatively with the measured $v_1(\eta), \, v_2(p_T)$ and $v_2(\eta)$. 
\end{abstract}
\maketitle

\section{Introduction}

The study of collective flow in relativistic nuclear collisions has potential to offer insights
into the equation of state of the produced matter \cite{whitepapers, collectiveflow}. Anisotropic 
flow is conveniently characterized by the Fourier coefficients \cite{methods}

\begin{equation}
v_n = \langle \cos n( \phi-\Psi_R ) \rangle
\end{equation}
where the angle brackets indicate an average over all the particles used, $\phi$ denotes the 
azimuthal angle of the outgoing particles, $\Psi_R$ is the orientation of the reaction plane, and 
$n$ denotes the harmonic.  The reaction plane is defined by the beam axis and the vector connecting 
the centers of the two colliding nuclei. The estimated reaction plane is called the event  plane, 
and its orientation is denoted above by $\Psi_R$.  The procedure used in the present study to 
estimate this angle is explained in Section II.C.  

Directed flow, $v_1$, is the first harmonic coefficient of the Fourier expansion of the 
final-state momentum-space azimuthal anisotropy, and it reflects the collective sidewards motion of 
the particles in the final state.   Both hydrodynamic and nuclear
transport models \cite{Hydro,Transport} indicate that directed flow  is a promising observable 
for investigating a possible phase transition, especially in the region of relatively low 
beam energy under investigation in the present paper \cite{bes, weight}.  In particular, the 
shape of $v_1$ as a function of rapidity, $y$, in the midrapidity region is of interest 
because it has been argued that it offers sensitivity to crucial details of the expansion 
of the participant matter during the early stages of the collision.  The models indicate that the 
evolving shape and orientation of the participant zone and its surface play a role in determining 
the azimuthal anisotropy measured among these particles in the final state. 

It has been known for a long time that in the general vicinity of beam rapidity, the directed 
flow is dominated by a ``bounce-off" effect, whereby particles are preferentially emitted within 
the reaction plane, and are collectively deflected towards the side that corresponds to being 
repelled from the participant zone \cite{Sorge, Herrmann}.  At RHIC energies, where the 
rapidity gap between beams is large (e.g., more than ten units at $\sqrt{s_{NN}}= 200$ GeV), the 
mid-rapidity region cannot be treated in terms of a monotonic interpolation between the rapidity 
regions on either side.  Over a region of $\sqrt{s_{NN}}$ spanning $\sim 20$ to 200 GeV, it is 
inferred that the slope $dv_1/d\eta$ (where $\eta$ is pseudorapidity) exhibits the algebraic sign 
associated with bounce-off near beam rapidities, but then $v_1(\eta)$ crosses zero at points well 
away from mid-rapidity, and the slope $dv_1/d\eta$ near mid-rapidity has the opposite sign 
\cite{PHOBOS_v1, v1-4systems}. 

The above phenomenon of $dv_1/d\eta$ or $dv_1/dy$ having opposite sign near mid-rapidity had
been predicted by hydrodynamic and nuclear transport models \cite{Csernai,Brachmann,Wiggle,Stocker} 
and variations on this theme had been given names like ``anti-flow", or ``third flow component" or 
``wiggle".  It has been argued that this is a possible signature of a phase transition between hadronic 
matter and quark gluon plasma, especially if it is observed for identified baryons \cite{Stocker}.  
However, it is also possible to explain the qualitative features of the sign reversal in $dv_1/dy$ 
in a purely hadronic picture by assuming strong but incomplete baryon stopping, together with 
strong space-momentum correlations caused by transverse radial expansion \cite{Wiggle}. 

Elliptic flow, $v_2$, is the second harmonic coefficient of the Fourier expansion.  The 
initial-state spatial eccentricity of the participant zone drives the process whereby the 
interactions produce an anisotropic distribution of momenta relative to the reaction plane.  The 
elliptic anisotropy saturates quite early in the collision evolution, although a little later than 
when directed flow saturates \cite{Sorge, Herrmann}.  Elliptic flow can provide information about 
the pressure gradients in a hydrodynamic description, and about the effective degrees of freedom, 
the extent of thermalization, and the equation of state of the matter created at early times.  
Studying the dependence of elliptic flow on system size, number of constituent quarks, and 
transverse momentum or transverse mass, are crucial to the understanding of the properties 
of the produced matter \cite{whitepapers}. 

 We find that directed flow violates the ``entropy-driven" multiplicity scaling which dominates all other soft observables. STAR has reported an intriguing new universal scaling of 
the phenomenon with collision centrality \cite{v1-4systems}.  Unlike the ratio of the elliptic flow parameter $v_{2}$ to the system initial eccentricity ($\epsilon$), which scales with
the particle density in the transverse plane, $v_{1}(\eta)$ at a given centrality is found to be independent of the system size, and varies only with the incident energy. The different scalings for 
$v_{2}/\epsilon$  and $v_{1}$ might arise from the way in which they are developed: to produce $v_{2}$, many momentum exchanges among particles must occur, while to produce
$v_{1}$, an important feature of the collision process is that different rapidity losses need to occur for particles at different distances from the center of the participant zone. This later quantity is 
related to incident energy.  A recent (3+1)-dimensional hydrodynamic calculation has successfully reproduced various aspects of the measured directed flow for Au+Au and Cu+Cu at 200 
GeV \cite{Bozek}. It is of interest to see if these features are followed at lower beam energies, as investigated in the current work.

 We report here the first measurements of the directed and elliptic flow in Cu+Cu collisions from the STAR~\cite{starnim} 
experiment at the Relativistic Heavy Ion Collider (RHIC) at $\sqrt{s_{NN}}=$ 22.4 GeV. The measurement 
of directed flow is presented as a function of pseudorapidity ($\eta$)  and elliptic flow is presented as
a function of pseudorapidity,  transverse momentum ($p_{T}$) and centrality.

The paper is organized as follows: Section II briefly describes the detectors used and provides  
details of the analysis methods.  In Section III, we present results on directed and elliptic flow 
($v_1$ and $v_2$) and compare these with models.  A summary is provided in Section IV.

\section{Experiment and Analysis}

\subsection{STAR detector subsystems}

The results presented here are based on data recorded by the STAR detector ~\cite{starnim} .  
The Time Projection Chamber (TPC)~\cite{startpc}  is the primary tracking device at STAR.  
It is 4.2 m long and 4 m in diameter and its acceptance covers $\pm 1.0 $ units of pseudorapidity and has full azimuthal coverage.  
The charged particle momenta are measured by reconstructing their trajectories  through the TPC. 

The charged particle reconstruction at forward rapidities is provided by STAR's Forward Time Projection Chambers (FTPCs)~\cite{starftpc}. 
The two  FTPCs are  located  around the beam axis, one at  each end of the STAR detector approximately 2.3 m from the nominal interaction region (IR),
 with acceptance in  the pseudorapidity range $2.5 \le |\eta| \le 4.0$, and with full azimuthal coverage. 

The Beam Beam Counter (BBC) detector subsystem ~\cite{BBC}  consists of two detectors mounted
around the beam pipe, each located outside the STAR magnet pole-tip at opposite ends
of the detector approximately 375 cm from the center of the nominal IR.  Each BBC detector
consists of nearly circular scintillator tiles arranged in four concentric rings
that provide full azimuthal coverage.  The two inner rings have a pseudorapidity coverage of 3.3 \lt $|\eta|$ \lt 5.0 and are used to reconstruct the first harmonic
event plane ($\Psi_{1}$),  for the directed flow analysis.  The outer BBC tiles were not used. 

\subsection{Experimental data sets}

The results presented in this paper  are from   Cu+Cu collisions at the nucleon-nucleon center of mass energy of 
$\sqrt{s_{NN}}$ = 22.4 GeV  with a minimum bias trigger based on BBC  coincidences~\cite{trigger,BBC}. We present the results 
for the 0-60\% most central events, for which the trigger efficiency was uniform. The primary collision vertex position along the beam 
direction $(V_z)$ has a broad Gaussian distribution with a root-mean-square  of 62 cm.  Only events within 30 cm of the center of the detector
are selected for this analysis. This value is chosen as a compromise between uniform detector performance within $|\eta| < 1.0$ and sufficient statistical 
significance of the measured observables. Event vertex is further required to be in the transverse direction within 2.0 cm from the center of the beam pipe in order to reject events which involve interactions with the beam pipe and  to minimize beam-gas interactions. For directed flow analysis we further apply an event cut  based on the maximum value of ADC signal from the BBC to avoid saturation of the BBC tiles. After these event cuts were applied the sample used for the directed flow analysis contained 350k events while the sample for the elliptic flow analysis consisted of  800k events.

Centrality classes in Cu+Cu collisions at $\sqrt{s_{NN}}$ = 22.4 GeV are defined using the number 
of charged particle tracks reconstructed in the TPC within pseudorapidity $|\eta| < 0.5 $ and passing within 3 cm of interaction  vertex.   The  uncorrected, i.e.  not corrected for acceptance and reconstruction efficiency,  multiplicity ($N_{\rm{ch}}$) distribution for events with a reconstructed primary vertex is shown in Fig. ~\ref{fig1}.  

\bef
\includegraphics[scale=0.45]{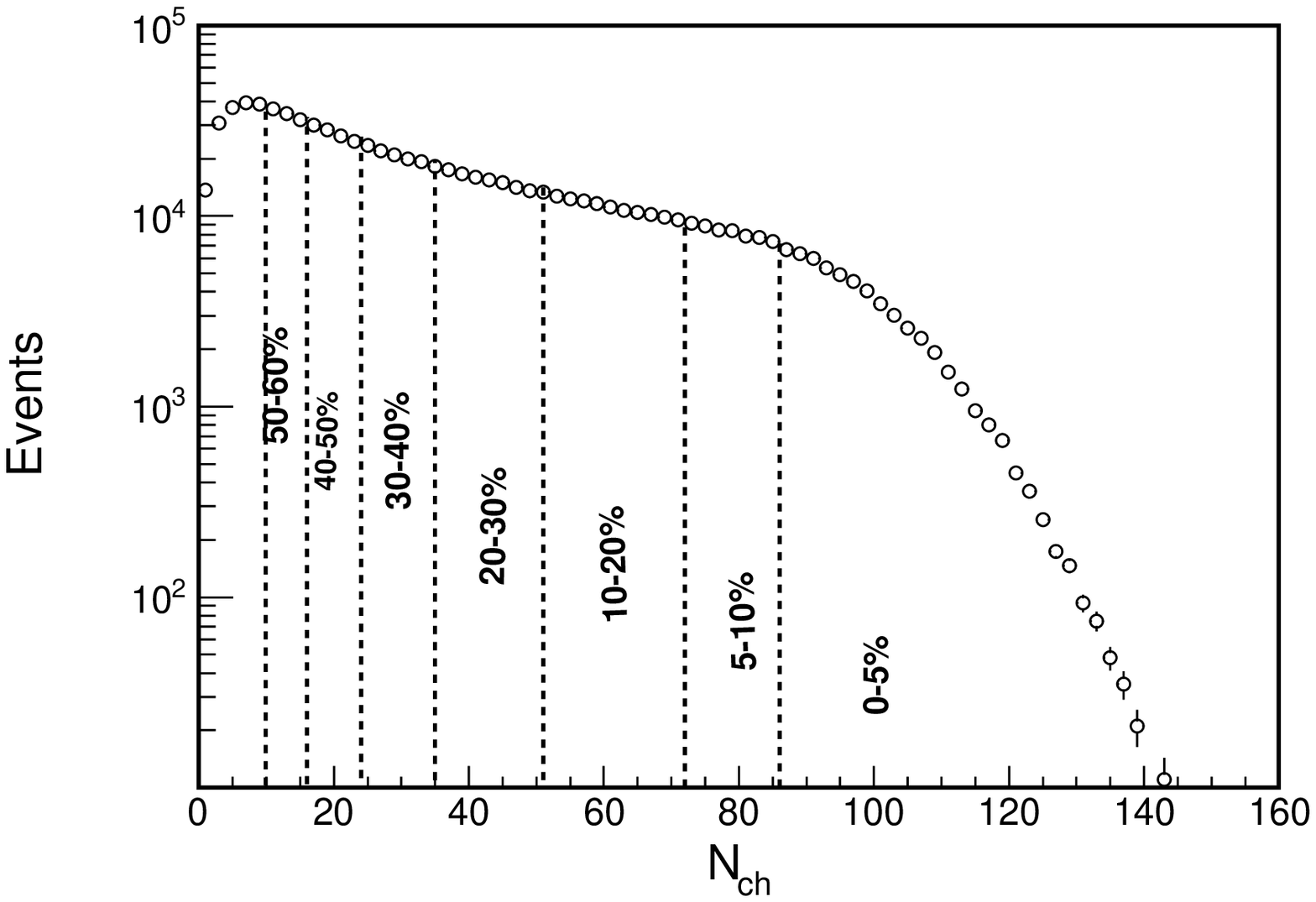}
\caption{ Uncorrected multiplicity distribution with $|\eta| < 0.5 $   in Cu+Cu collisions at $\sqrt{s_{NN}}$ = 22.4 GeV. 
          Events with $ N_{ch} > 10$ are selected for the present analysis.}
\label{fig1}
\eef
  
The directed and elliptic flow analyses were carried out on tracks that had transverse momenta $0.1 < p_{T} < 4.0$ GeV$/c$, passed within 
3 cm of the primary vertex, had at least 15 space points in the main TPC acceptance  $(|\eta| \lt 1.0)$ or 5 space points in the case of tracks in
 the FTPC acceptance ($2.5 \lt |\eta| \lt 4.0$) and the ratio of the number of actual space points to the maximum possible number of space points for that
 track’s trajectory was greater than 0.52.
 
\subsection{Flow methods}

For any Fourier harmonic \emph{n},  the event flow vector ($Q_n$) and the event plane angle ($\Psi_n$) are defined by~\cite{methods}

\begin{equation}  
Q_n \cos (n\Psi_n)  = Q_{nx} = \sum_i w_i\cos (n\phi_i),
\end{equation}
\begin{equation} 
Q_n \sin (n\Psi_n) = Q_{ny} = \sum_i w_i\sin (n\phi_i), 
\end{equation}
\begin{equation} 
\Psi_n  =  \left( \tan^{-1} \frac{Q_{ny}}{Q_{nx}} \right)/n,
\end{equation}
where sums extend over all particles $i$ used in the event plane calculation, and $\phi_i$ and $w_i$ are the laboratory azimuthal angle and the weight for 
the $i$-th particle, respectively.  In the case where  we construct the event plane using tracks reconstructed from hits in the TPC or FTPC, weight for each particle 
is taken to be $p_{T}$ in GeV/$c$  up to 2.0  GeV/c and constant at 2.0 GeV/c for higher $ p_{T}$~\cite{weight}.  In this case we denote the resulting  flow harmonic as 
$ v_{n} \{\textrm{TPC}\} $ or $ v_{n} \{\textrm{FTPC}\} $.  In those cases where the event plane is constructed using particle trajectories determined from hits in the BBC
detectors,  $\phi_i$ denotes the fixed azimuthal angle of the center of  the $i$th BBC tile, and $w_i$ is the energy deposition (the ADC signal, $A_i$) in the 
$i$th BBC tile and  $w_i$ is  calculated from the ADC signals $A_i$, where 

\begin{equation}
 w_i = \frac{A_{i}}{\sum A_{i}}.
\end{equation}
 The corresponding flow harmonic is here denoted as $v_{n} \{\textrm{BBC}\}$.  In all cases, the flow harmonic or
coefficient is calculated by

\begin{equation} \label{v2EP2} v_{n}\ =\
\frac{\langle \cos n(\phi-\Psi_n) \rangle}{\langle \cos n(\Psi_n-\Psi_R) \rangle} . 
\end{equation}
 where $\Psi_R$ is the true reaction plane angle, $\Psi_n$ is the event plane and  the angle brackets here denote an average over all the particles in a specific bin in centrality. Tracks used 
for the calculation of $v_n$ are excluded from the calculation of the event plane to remove 
self-correlation. The finite number of tracks limits the angular resolution of the
reconstructed event plane. Consequently, the flow coefficient that is obtained with
the reconstructed event plane, i.e.  the numerator in equation (6), requires an event plane resolution correction~\cite{methods}, 
which is the denominator in equation (6). The event plane resolution is estimated from the correlation of 
the event planes of two sub-events \cite{methods}. 
\bef
\includegraphics[scale=0.45]{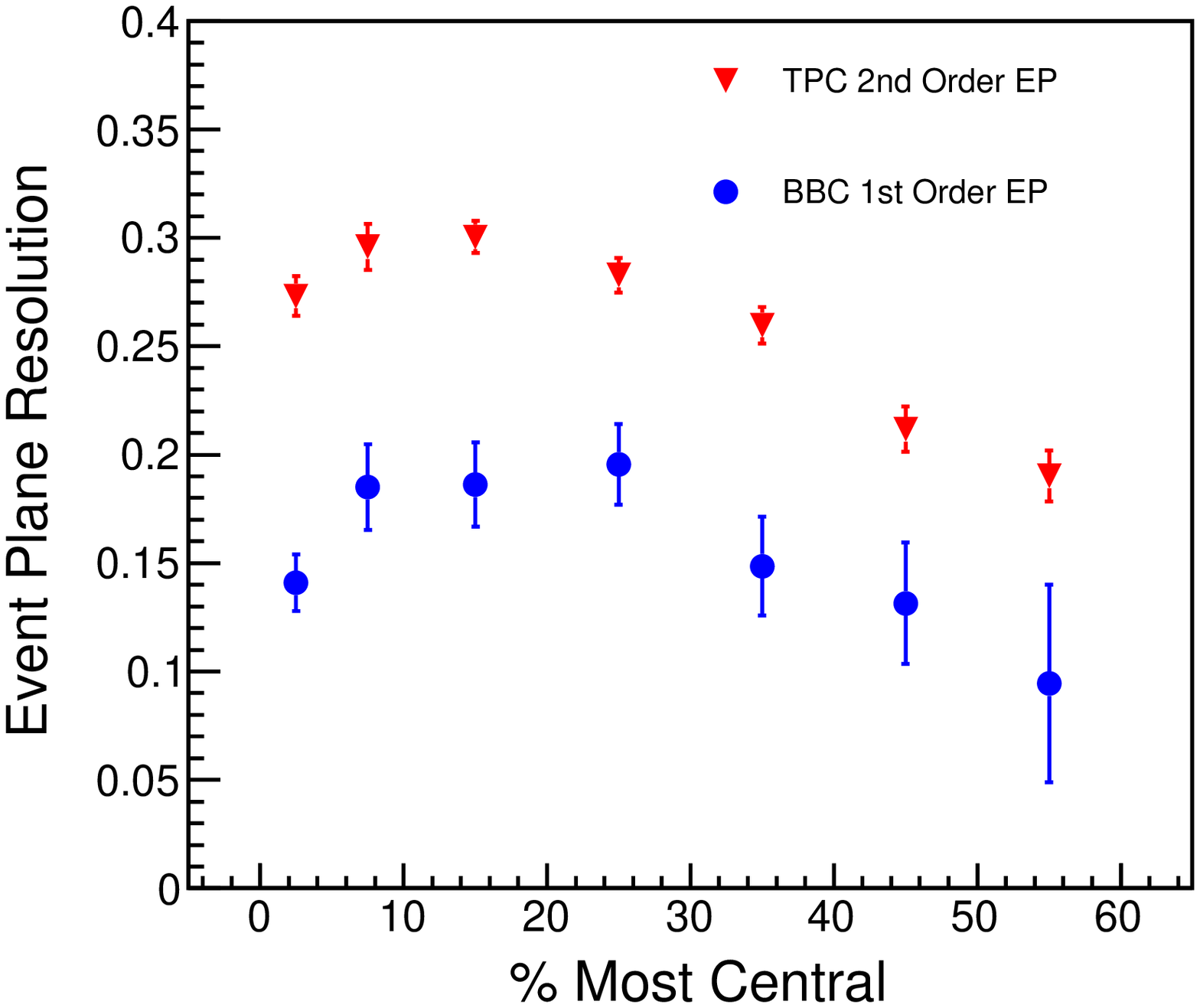}
\caption{ The event plane resolution
 measured using the TPC (second order) and using the BBC (first order)  are shown as
a function of collision centrality for Cu+Cu collisions at $\sqrt{s_{NN}}$ = 22.4 GeV.
Errors are statistical only.}
\label{fig2}
\eef 
 In the case of elliptic flow analysis, three different ways 
have been tested for choosing the sub-events: 
(1) Particles with pseudorapidity $-1.0 \lt \eta \lt -0.3$ are assigned to one sub-event and 
particles with $0.3 \lt \eta \lt 1.0 $ to the other sub-event.  The gap between the two 
pseudorapidity regions ensures that short-range correlations such as Bose-Einstein 
interference or Coulomb final-state interactions contribute negligibly to the observed 
correlations \cite{flow1},  and also may reduce possible jet effects, such as the minijet interpretation discussed in Ref.\cite{minijet}.  (2) Particles are assigned randomly to two sub-events. (3) Positive particles are  assigned to one sub-event and negative particles to the other. Figure ~\ref{fig2} shows the event plane resolution as a function of centrality, where the 
TPC second-order event plane resolution was determined using the first (pseudorapidity) method described above. The average event plane resolution for the \emph{n} = 2 plane for $v_{2} \{\textrm{TPC}\}$ is  $0.26 \pm 0.01$ for collisions with  centrality 0-60 \%.

The BBC event plane  obtained from one BBC detector is called a sub-event plane.  A combination of the sub-event plane vectors for both BBC
detectors provides the full event plane.  In the $v_1\{\textrm{BBC} \}$ method, we used the BBC full event 
plane to obtain directed flow in the TPC pseudorapidity range ($|\eta| < 1.0 $).  A self-correlation arises if $v_1$ is obtained using particles from 
the same pseudorapidity region as used for the event plane reconstruction.  This problem can arise in the $v_1\{\textrm{BBC}\}$ analysis, because there is 
partial overlap in pseudorapidity coverage between the FTPC and the BBC.  To avoid this, when $v_{1}$   was obtained in the  FTPC coverage  $-4.0 < \eta <  -2.5 $, 
the event plane was constructed using the BBC  covering  $3.3 < \eta < 5.0$, and conversely, when using the FTPC coverage  $2.5 < \eta < 4.0$ to determine $v_{1} $,
the event plane was determined with the  BBC covering  $-5.0< \eta < -3.3$.  The first order event plane resolution as a function of centrality is shown in Fig.~\ref{fig2}. The average event plane resolution for the \emph{n} = 1 plane for $v_{1} \{\textrm{BBC}\}$ is  $0.16 \pm 0.03$ for collisions with  centrality 0-60\%.

\section{Results}

\subsection {Directed Flow Results}
\bef
\begin{center}
\includegraphics[scale=0.45]{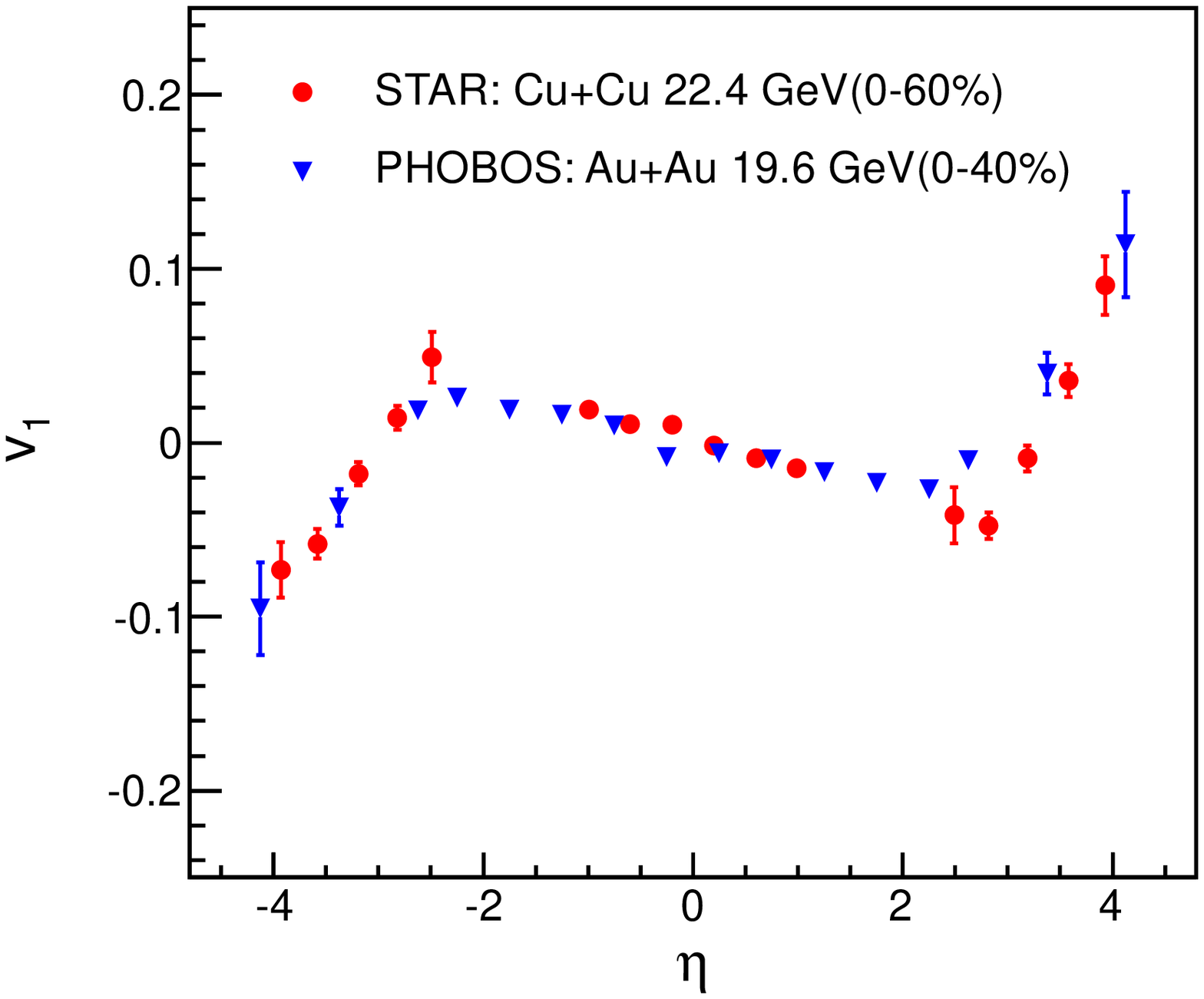}
\caption{ Charged hadron $ v_{1} \{\textrm{BBC}\}$ vs. $\eta$ for 0--60\% centrality Cu+Cu collisions at $\sqrt{s_{NN}} 
= 22.4$ GeV. The errors shown are statistical. Systematic errors are discussed in Section III.C. 
 Results are compared to $v_1$  from 0--40\% centrality Au+Au collisions at
$\sqrt{s_{NN}} = 19.6$ GeV from the PHOBOS collaboration~\cite{PHOBOS_v1}. }
\label{fig3}
\end{center}
\eef
Figure~\ref{fig3} shows charged hadron $v_1\{{\textrm {BBC}}\}$ in Cu+Cu collisions for 0--60\% 
centrality at $\sqrt{s_{NN}} = 22.4$ GeV as a function of  $\eta$, compared to that for  0-40\% central Au+Au collisions 
at $\sqrt{s_{NN}} = 19.6$ GeV measured by the PHOBOS experiment~\cite{PHOBOS_v1}.  The PHOBOS results are quite similar, notwithstanding the difference in 
system size, and the fact that the centrality range and beam energy are not the same.  
At 200 GeV and 62.4 GeV, we have previously reported that directed flow is not different  within 
errors  for Au+Au and Cu+Cu \cite{v1-4systems}.  We find that  
this behavior extends to lower energies. Directed flow provides information about the collision process that complements the more widely studied elliptic flow. Elliptic flow is
developed after a number of momentum exchanges among particles, and the number of
such exchanges depends on the dimensions of the participant system and on its
density.  Consequently, for a given collision centrality, elliptic flow varies with
the mass of the colliding nuclei.  In contrast, the observation that directed flow
does not vary with the mass of the colliding nuclei is a reflection of the 
different mechanism that generates $v_1$: here, the relevant feature is the rapidity
shift undergone by particles that are initially located at different distances from
the center of the participant volume \cite{Wiggle} - a fundamental characteristic of
the relativistic heavy-ion interaction process.  

\bef
\begin{center}
\includegraphics[scale=0.45]{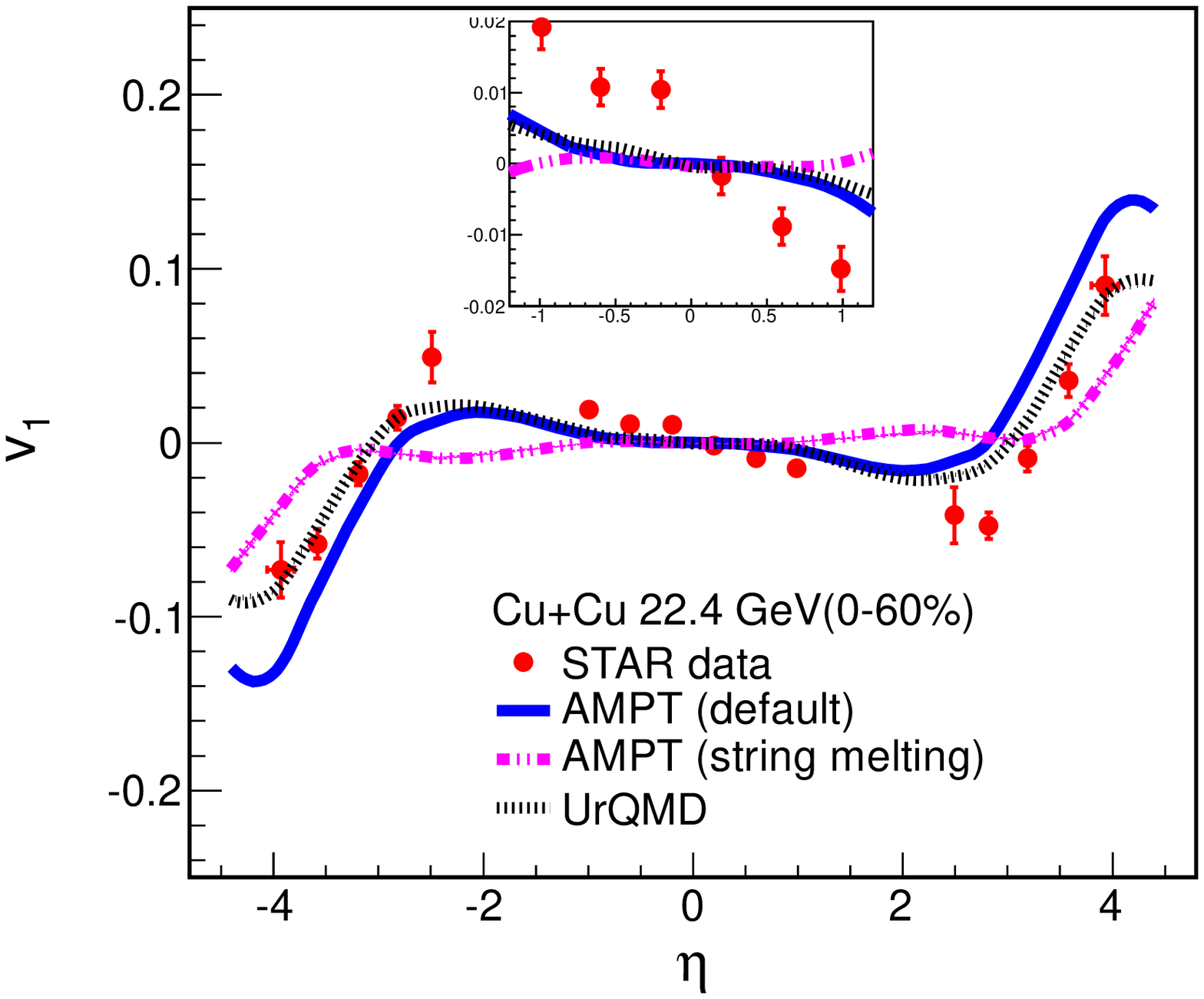}
\caption{ Comparison of the measured $ v_{1} \{\textrm{BBC}\}$  as a function of $\eta$  in 0-60\% Cu+Cu
  collisions at $\sqrt{s_{NN}} = 22.4$  GeV with model predictions. The inset
shows the central $\eta$ region in more detail.  The errors are statistical only.}
\label{fig4}
\end{center}
\eef

In Fig.~\ref{fig4} we compare our measurements to the results of the A Multi Phase Transport (AMPT) \cite{ampt} and Ultra Relativistic Quantum Molecular Dynamics (UrQMD) \cite{urqmd} models. Around
midrapidity, the models predict substantially smaller slope of $v_{1}(\eta)$ than observed in the data,  whereas at forward rapidities, the models differ among themselves and bracket the data.  The fact that  the tested models do not reproduce the observed pattern of $v_1$ as a function of  pseudorapidity   implies the need for further evolution in the model descriptions.  

\bef
\begin{center}
\includegraphics[scale=0.45]{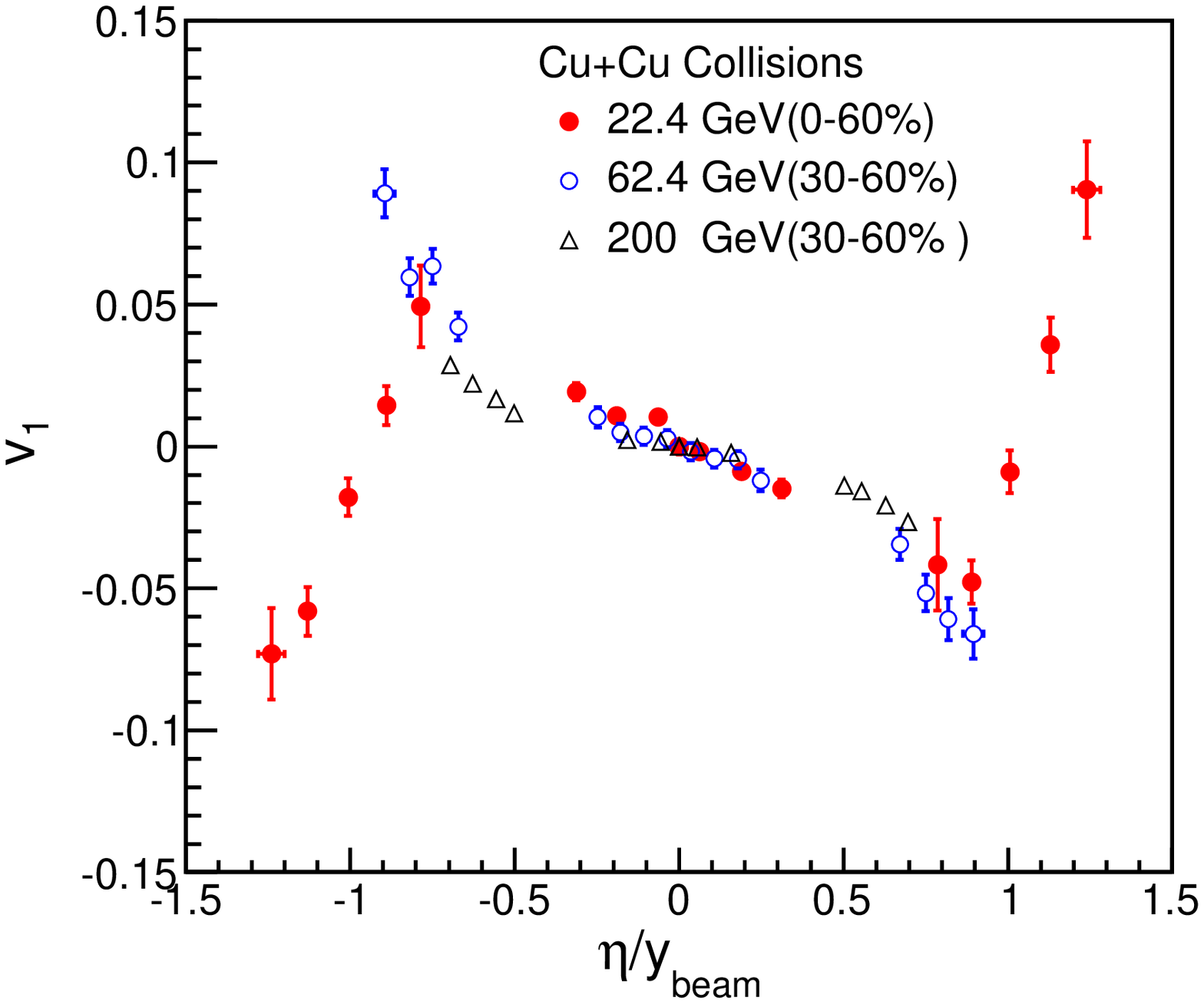}
\caption{Charged hadron $v_1$ as a function of $\eta$, scaled by the respective $y_{\rm{beam}}$  
for the three beam energies 22.4, 62.4 and 200 GeV.   The results for 62.4 and 200 GeV are  
for $30-60$\% centrality Cu+Cu collisions previously reported by STAR~\cite{v1-4systems}. For 22.4 GeV, 
the plotted results are for $0-60$\% centrality. }
\label{fig5}
\end{center}
\eef

Figure \ref{fig5} shows charged hadron $v_1$ as a function of pseudorapidity 
scaled by the respective beam rapidity ($y_{\rm beam}$) values for the three 
beam energies 22.4, 62.4 and 200 GeV in Cu+Cu collisions \cite{v1-4systems}.  
For Au+Au collisions over a range of $\sqrt{s_{NN}}$ spanning 19.6 to 200 
GeV, it is an empirical observation that $v_1(\eta/y_{\rm beam})$ lies close 
to a single common curve for all beam energies \cite{PHOBOS_v1, v1-4systems}, 
and this type of scaling was first observed by NA49 at the SPS \cite{NA49_v1}.  
The new results reported here for Cu+Cu at 22.4 GeV extend the range of 
applicability of this scaling behavior. 
 
\bef
\begin{center}
\includegraphics[scale=0.45]{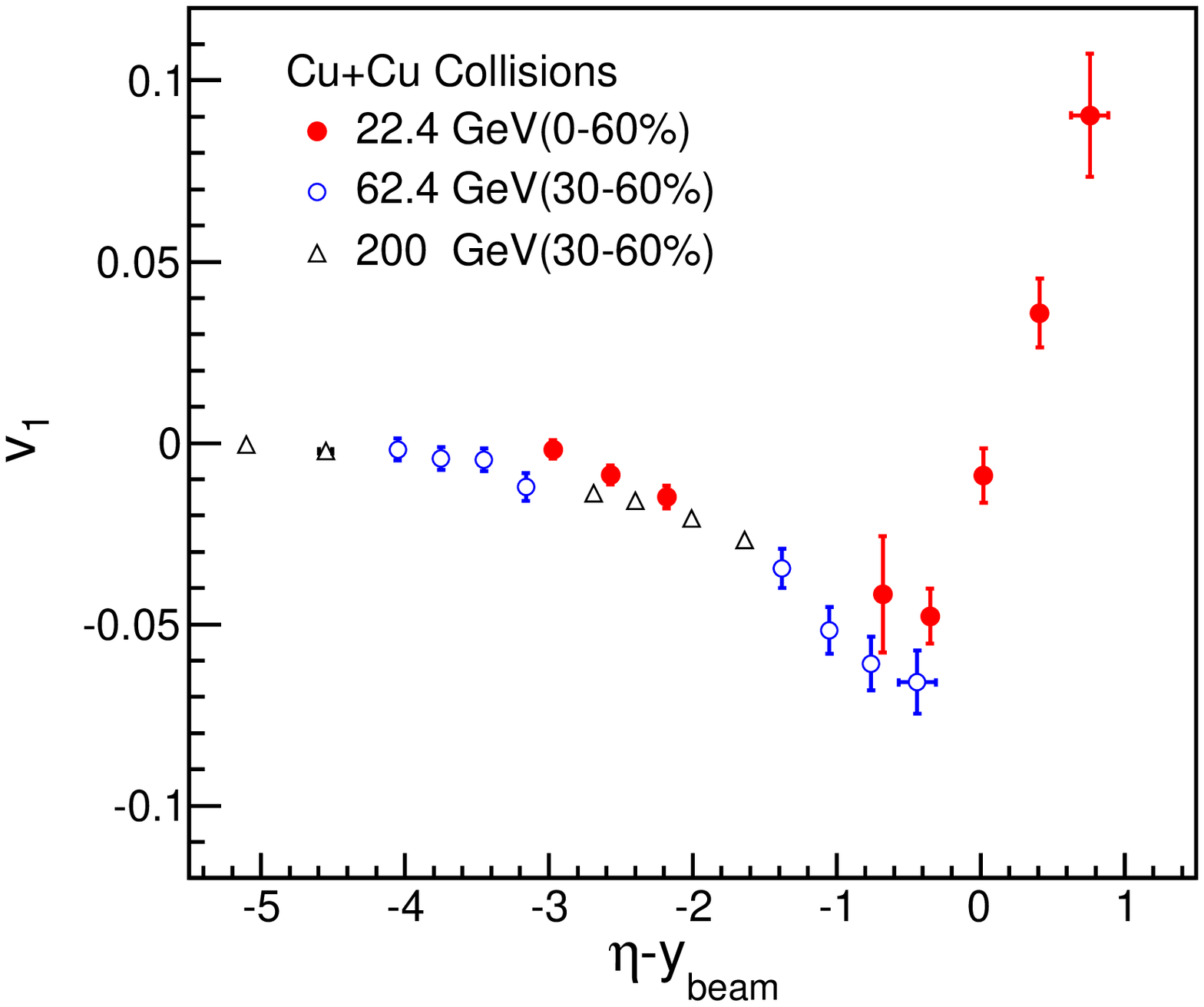}
\caption{Charged hadron $v_1$ as a function of $\eta$ - $y_{\rm{beam}}$ values 
for the three beam energies 22.4, 62.4 and 200 GeV.  The results for 62.4 and 200 GeV are 
for 30--60\% centrality Cu+Cu collisions previously reported by STAR~\cite{v1-4systems}.}
\label{fig6}
\end{center}
\eef

Figure \ref{fig6} shows charged hadron $v_1$ as a function of $\eta$ -  $y_{\rm{beam}}$, i.e. in the projectile frame for three beam energies  22.4, 62.4 and 200 GeV \cite{v1-4systems}.  The data support the limiting fragmentation hypothesis~\cite{v1-4systems} in the region $-2.6  \lt \eta-y_{\rm{beam}}\lt 0 $.
\subsection{Elliptic Flow Results}
\bef
\begin{center}
\includegraphics[scale=0.45]{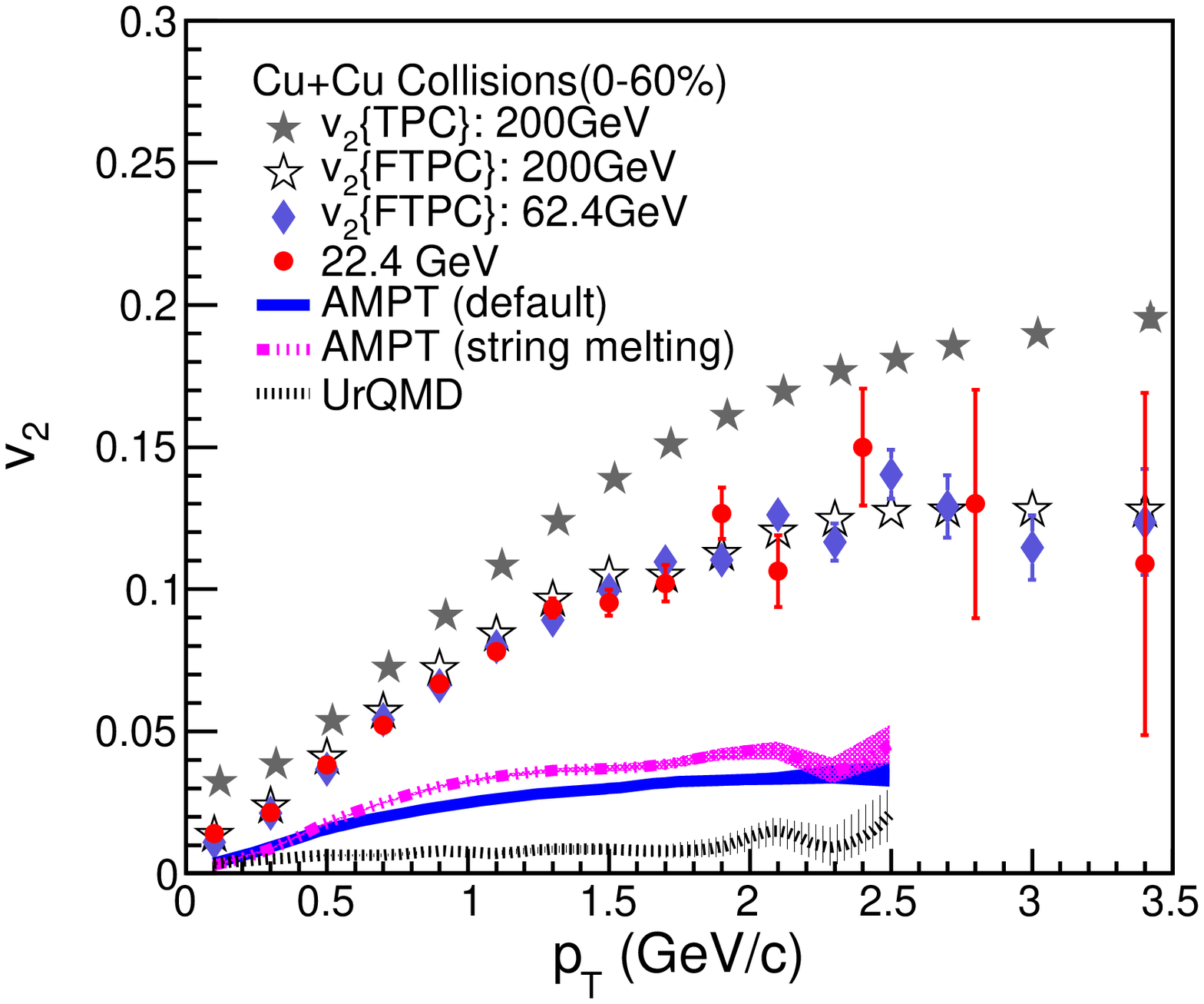}
\caption{Elliptic flow versus $p_T$ for charged hadrons from Cu+Cu collisions at 0--60\% centrality at
$\sqrt{s_{NN}} = 22.4$ GeV  measured with the  subevent method with a pseudorapidity gap of 0.3 units compared with previously published STAR results for 200 and 62.4 GeV Cu+Cu \cite{CuCuPaper} measured with full TPC event plane method $v_{2} \{\textrm{TPC}\}$ and full FTPC event plane method $v_{2} \{\textrm{FTPC}\}$.  The error bars are  statistical. Results are also compared to $v_{2} (p_{T} )$  model calculations.} 
\label{fig7}
\end{center}
\eef

Figure~\ref{fig7} shows $v_{2}(p_T)$ for charged hadrons from Cu+Cu collisions at $\sqrt{s_{NN}} = 22.4$ GeV measured with the
subevent method with a pseudorapidity gap of 0.3 units.  Also shown are the previously published STAR results for 200 and 62.4 GeV Cu+Cu \cite{CuCuPaper} measured with the full TPC 
event plane method $v_{2} \{\textrm{TPC}\}$ and full FTPC event plane method $v_{2} \{\textrm{FTPC}\}$.  The significant observed difference between $v_2 \{\textrm{TPC}\}$ and
 $v_2 \{\textrm{FTPC}\}$ at 200 GeV is discussed in detail in Ref.~\cite{CuCuPaper} and is attributed to the relatively large 
non-flow present in $v_2 \{\textrm{TPC}\}$ for Cu+Cu at the top RHIC energy.
We observe that the elliptic flow at 22.4 GeV is systematically lower than $v_{2} \{\textrm{TPC}\}$ at 200 GeV, however, it is similar to $v_{2} \{\textrm{FTPC}\}$ at  200 and 62.4 GeV, consistent 
with the earlier observation \cite{vtwodiff}.  For comparison, we also  show $v_{2} (p_{T})$ from the UrQMD and AMPT models.  The models do not agree with the data, but they do show an 
increase in $v_{2}$ with transverse momentum similar to the data and plateau at much lower values of $ p_{T}$. The small sample size in the present analysis precludes an extension of the 
measurements  to identified particle $v_{2} $.   
\bef
\begin{center}
\includegraphics[scale=0.45]{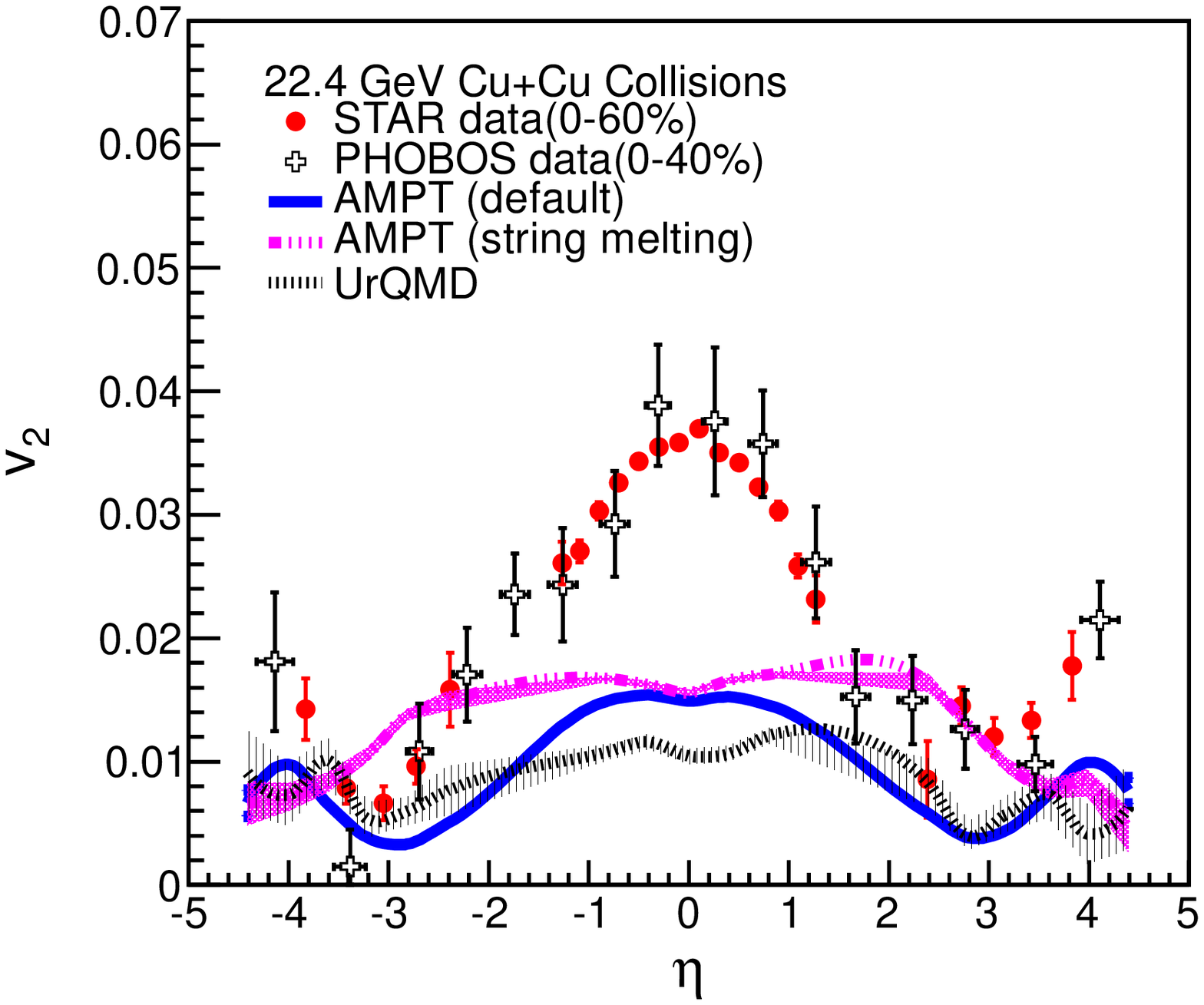}
\caption{Elliptic flow $v_2(\eta)$ for charged hadrons from Cu+Cu collisions at 0--60\% centrality at $\sqrt{s_{NN}} = 22.4$ GeV.  The present STAR results
are compared to the measurement from the  PHOBOS \cite{PH0BOS_cu22} collaboration for Cu+Cu at 22.4 GeV. The PHOBOS results include statistical and systematic errors whereas the STAR results
are plotted with statistical uncertainties only, and systematic errors are discussed in Section III C. Results are also compared to $v_2(\eta)$ calculations from the indicated models.}
\label{fig8}
\end{center}
\eef

Figure~\ref{fig8} shows $v_2(\eta)$ for charged hadrons from Cu+Cu collisions at 0-60\%  centrality at  $\sqrt{s_{NN}} = 22.4$ GeV.  These STAR results are compared to 
published measurements from the PHOBOS Collaboration for  0-40\% central collisions at  $\sqrt{s_{NN}} = 22.4$ GeV  \cite{PH0BOS_cu22}. The PHOBOS  error bars include statistical and systematic errors, 
whereas the STAR data are plotted with statistical error bars only. Systematic errors are discussed in Section III.C.  The STAR results for $v_2(\eta)$ are consistent within errors with the PHOBOS 
data.  We also  compare with corresponding predictions from the AMPT and UrQMD models. These models  underpredict the data at midrapidity, but do show a trend that is similar to the data for $ |\eta|> 2.0 $.  

\bef
\begin{center}
\includegraphics[scale=0.45]{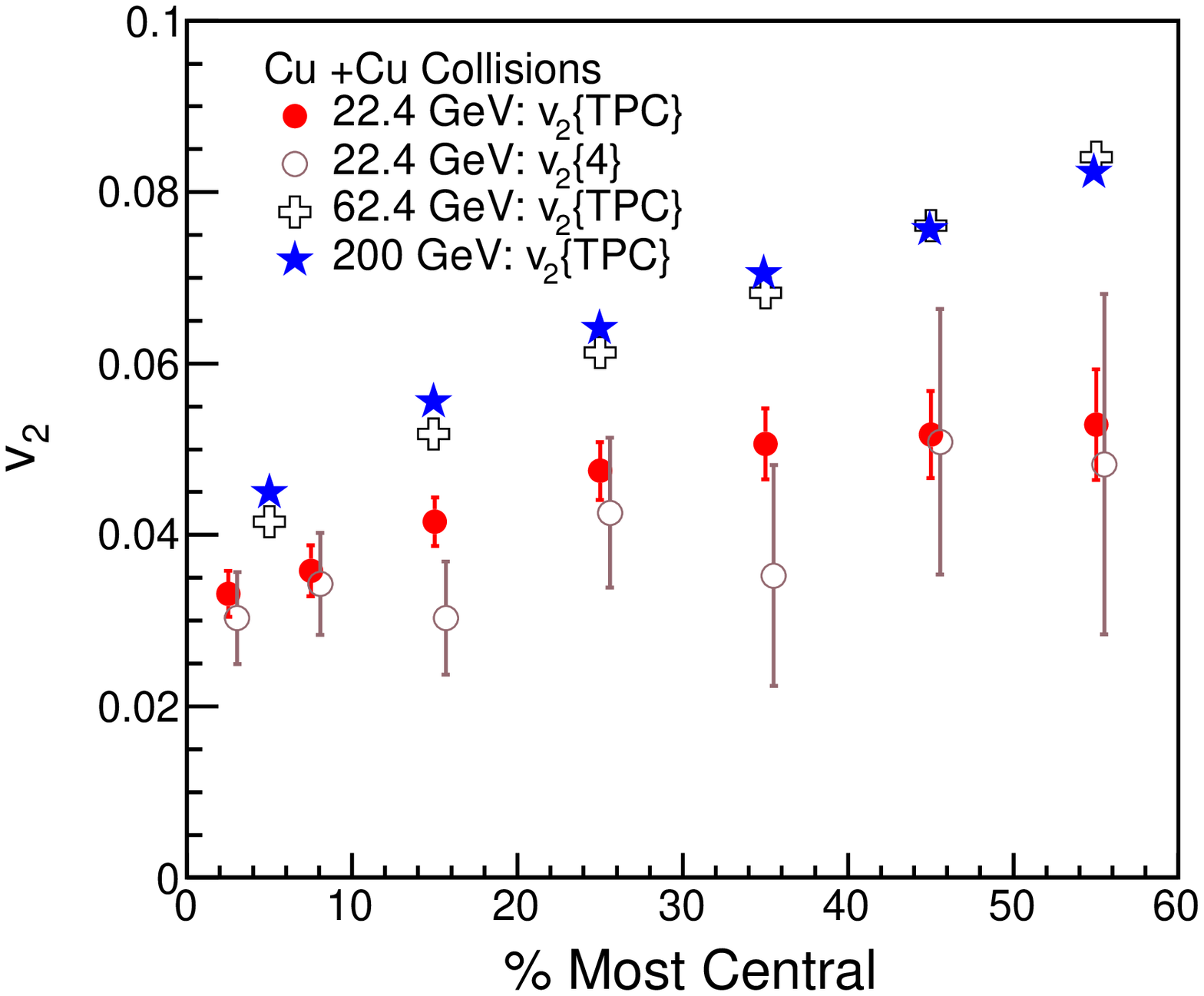}
\caption {Elliptic flow $v_2\{\textrm{TPC}\}$ and $v_2\{4\}$  as a function of centrality for charged hadrons from Cu+Cu collisions at $0-60\%$ centrality at $\sqrt{s_{NN}} = 22.4$ GeV, compared with previously published results from the STAR collaboration at 200 GeV and 62.4 GeV. }
\label{fig9}
\end{center}
\eef

Figure~\ref{fig9} presents $v_2 \{\textrm{TPC}\}$ for $0.1 <  p_T < 2.0 $ GeV/$c$ and $|\eta|<1.0 $, as a function of centrality for charged hadrons from Cu+Cu collisions at 0--60\% centrality at $\sqrt{s_{NN}} = 22.4$ GeV, plotted along with previously published results from the STAR collaboration at 200 GeV and 62.4 GeV \cite{CuCuPaper}. 
The present $v_2\{\textrm{TPC}\}$ result was obtained using $\eta$ sub-events \cite{flow1} with a gap of 0.3 units in pseudorapidity while the published results are based on the 
full TPC. Also shown in Fig.~\ref{fig9} is a four-particle cumulant \cite{4ptcumulant} analysis $v_2\{ 4 \}$. This analysis method helps reduce some types of systematic error (see below) but 
generally requires more statistics than a method like $v_2\{\textrm{TPC}\}$. The statistical errors on our $v_2 \{ 4 \}$ measurements are small enough to be useful in the case of the more central 
collisions. The results from this $v_2 \{4\}$  analysis  agree, within statistical errors, with the $v_2 \{\rm{TPC}\} $ results, although there are hints of $v_2 \{4\}$ being systematically lower than 
$v_2 \{\rm{TPC}\}$ for the more central data points. The beam energy dependence of the $p_T$-integrated $v_2$ mainly comes from the energy dependence of the mean $p_T$ and the 
difference between the event plane reconstruction with and without a pseudorapidity gap. 

\subsection{Systematic uncertainties}

Any analysis of collective flow needs to consider the possible systematic uncertainty arising 
from non-flow \cite{methods}, which refers to azimuthal correlations not related to 
the reaction plane orientation. Non-flow can arise from resonances, jets, strings, quantum 
statistics effects, final state interactions (particularly Coulomb effects), and momentum 
conservation.  Different methods used to measure anisotropic flow are affected by non-flow 
in different ways, and are used in this analysis to guide our estimates of the systematic 
uncertainty.  In general, non-flow arising from jets can be expected 
to be less troublesome at lower beam energies. Moreover, as described
earlier, the relatively large pseudorapidity gap between the STAR TPC and the BBCs is
helpful in suppressing non-flow.  
   
When the event plane is determined from a detector that is not symmetric around $\eta =0$, we need to account for correlations in the measured directed flow due to momentum conservation \cite{momentum}.  The desired BBC $\eta$ symmetry is present for our $v_1$ analysis in the $\eta$ region of the central TPC, but is a source of possible concern for the FTPC $\eta$ region. The overlap in $\eta$ acceptance between the BBC and FTPC is only partial, and therefore it is feasible to compare $v_1\{\textrm{BBC}\}_{\rm full}$ (where both east and west BBCs are used),   
with $v_1 \{\textrm{BBC}\}_{\rm sub}$, using either the east or west BBC event plane for $2.5 < |\eta| < 3.3$.  We find that the difference is less than 10\%, and an extrapolated average correction has been applied to $v_1\{\textrm{BBC}\}$ for $|\eta| > 3.3$.  

The measured $v_1$ must be antisymmetric about mid-pseudorapidity within statistical errors.  Any larger difference is due to systematic errors.  Previous detailed studies point to the maximum forward-backward difference as a viable estimate of the overall systematic uncertainty when the pseudorapidity gap is large \cite{flow1}.  We conclude that the overall systematic uncertainty in our determination of $v_1$ is approximately 15\% in the FTPC region, and about 10\% in the central TPC region.  

In our analysis of elliptic flow, $v_2$, unlike for directed flow, $v_1$, we do not have the advantage of a wide $\eta$ gap to help ensure that non-flow background effects are minimized. 
To study possible systematic effects associated with short range non-flow correlations, a four-particle cumulant \cite{4ptcumulant} analysis $v_2\{ 4 \}$ as a function of centrality has been investigated, and is plotted in Fig.~\ref{fig9}. This method suppresses non-flow correlations involving fewer than four particles. The statistical errors on our $v_2 \{ 4 \}$ measurements are small enough to be useful for the most central collisions, but grow to a few tens of percent at the other end of our studied centrality range. The $v_2 \{\textrm{TPC}\}$ and $v_{2}\{4\}$ measurements agree within statistical errors, although the observed systematic difference, which might arise from non-flow effects, amounts to about 9\% for the $0-10\%$ most central collisions. 

In the elliptic flow measurement, we have used the $\eta$ sub-event method with a gap of 0.3 units in pseudorapidity ($\eta$). As noted in Section II.C, such a gap suppresses short-range correlations such as Bose-Einstein interference and Coulomb final-state interactions.  To estimate the non-flow contributions to our estimate of $v_2\{\textrm{TPC}\}$ due to these short-range correlations, we have studied variations in the resulting $v_2 \{\textrm{TPC}\}$ induced by varying the event vertex selection  
along the beam direction, by varying the DCA cut value, and by varying the size of the pseudorapidity gap between the sub-events in the $\eta$ sub-event method. Tests of this type suggest that the systematic error on $v_2$ is on the order of 10\%.  However, based on an alternative approach \cite{minijet} to fitting and interpreting the measured data at 62.4 GeV and 200 GeV, it is argued that the systematic uncertainty could be larger.  
  
\section{Summary}

In this study of Cu+Cu collisions at $\sqrt{s_{NN}} = 22.4$ GeV, we present results at midrapidity and at forward rapidity for directed flow $v_1$ and elliptic flow $v_2$ as a function of pseudorapidity, and for elliptic flow $v_2$ as a function of transverse momentum at midrapidity for charged hadrons.  For our directed flow measurement, non-flow correlations are expected to be strongly suppressed by the large pseudorapidity gap between the detector used for event plane determination (BBC) and the main tracking detector (TPC) used in our directed flow measurement. Our estimate of the systematic error on the directed flow $v_1 \{\textrm{BBC}\}$  is not more than 10\% to 15\% for the TPC and FTPC regions, respectively. Our findings for Cu+Cu and Au+Au at 22.4 GeV extend observations previously made at 62.4 GeV and 200 GeV that directed flow is independent of system size in this energy  region. Our findings also demonstrate that $v_1 (\eta/y_{\rm beam})$ remains independent of beam energy down to 22.4 GeV, a scaling behavior that has already been established at 62.4 GeV and 200 GeV.  We find that directed flow violates the ``entropy-driven" multiplicity scaling which dominates all other soft observables.  An important feature associated with generation of $v_{1}$  in the of the collision process is that different rapidity losses need to occur for particles at different distances from the center of the participant zone, which is beam-energy dependent. Measurements of the elliptic flow for 22.4 GeV Cu+Cu collisions are also presented. We compare $p_T$-integrated $v_2$ with measurements at higher energies. The $p_T$-dependence of the measured $v_2$ at 22.4 GeV is similar to that at 62.4 and 200 GeV. UrQMD and AMPT models (the latter both with and without string melting) do not agree with the present measurements for both first $(v_{1})$ and second $(v_{2})$ coefficients. 

{\center \bf Acknowledgments \center }
We thank the RHIC Operations Group and RCF at BNL, the NERSC Center at LBNL and the Open Science Grid consortium for providing resources and support. This work was supported in part 
by the Offices of NP and HEP within the U.S. DOE Office of Science, the U.S. NSF, the Sloan Foundation, the DFG cluster of excellence `Origin and Structure of the Universe' of Germany, CNRS/
IN2P3, FAPESP CNPq of Brazil, Ministry of Ed. and Sci. of the Russian Federation, NNSFC, CAS, MoST, and MoE of China, GA and MSMT of the Czech Republic, FOM and NWO of the 
Netherlands, DAE, DST, and CSIR of India, Polish Ministry of Sci. and Higher Ed., Korea Research Foundation, Ministry of Sci., Ed. and Sports of the Rep. Of Croatia, and RosAtom of Russia.

\end{document}